\documentclass[aps,showpacs,preprintnumbers,amsmath,amssymb]{revtex4}

\UseRawInputEncoding

\usepackage{graphicx}
\usepackage{subfig}
\usepackage{esint}

\usepackage{amsmath,amssymb,amsfonts}

\begin{document}
\baselineskip=0.8 cm
\captionsetup{justification=raggedright,singlelinecheck=false}
\title{Polarized image of a rotating black hole in Scalar-Tensor-Vector-Gravity theory}

\author{Xin Qin$^{1}$,
Songbai Chen$^{1,2}$\footnote{Corresponding author: csb3752@hunnu.edu.cn}, Zelin Zhang$^{1}$,
Jiliang Jing$^{1,2}$ \footnote{jljing@hunnu.edu.cn}}
\affiliation{ $ ^1$ Department of Physics, Key Laboratory of Low Dimensional Quantum Structures
and Quantum Control of Ministry of Education, Synergetic Innovation Center for Quantum Effects and Applications, Hunan
Normal University,  Changsha, Hunan 410081, People's Republic of China
\\
$ ^2$Center for Gravitation and Cosmology, College of Physical Science and Technology, Yangzhou University, Yangzhou 225009, People's Republic of China}

\begin{abstract}
\baselineskip=0.6 cm
\begin{center}
{\bf Abstract}
\end{center}
The polarized images of a synchrotron emitting ring are studied in the spacetime of a rotating black hole in the Scalar-Tensor-Vector-Gravity  (STVG) theory. The black hole owns an additional dimensionless MOG parameter described its deviation from Kerr black hole.
The effects of the MOG parameter on the observed polarization vector and Strokes $Q-U$ loops depend heavily on the spin parameter, the magnetic field configuration, the fluid velocity and the observation inclination angle. For the fixed MOG parameter, the changes of the polarization vector in the image plane are similar to those in the Kerr black hole case. The comparison of the polarization images between Kerr-MOG black hole and M87* implies that there remains some possibility for the STVG-MOG theory.

\end{abstract}

 \maketitle
\newpage
\section{Introduction}

The black hole images of M87* \cite{EHT1,EHT2,EHT3,EHT4,EHT5,EHT6} and Sgr A*\cite{EHT7}, photographed by the Event Horizon Telescope (EHT), strongly confirm the existence of black holes in our Universe, which have greatly stimulated the study of
black hole physics, both theoretically and experimentally. These images achieving a diffraction limited angular resolution bring us a lot of information from the strong field region near black hole. In the first polarized images of the black hole M87* \cite{EHT8,EHT9}, the twisting polarization pattern with prominent rotationally symmetric mode reveals that there is a strong magnetic field
around black hole in terms of electrons synchrotron radiations.
Moreover, the analysis  show that the polarization patterns depend also on the strongly curved
spacetime near the black holes. Thus, study of polarized images of black holes is beneficial to probe the matter distribution and the related  physical process around black holes \cite{PZ1,PZ2,PZ3,PZ4,PZ5,PZ6,PZ7,PZ8,PZJG1,PZJG2,PZJG3,PZJG4,PZJG5,PZJG6,PZJG7,PZJG8,PZJG9,PZJG10,PZJG11}, even check theories of gravity.

Exact description for polarized images of black holes must resort to numerical simulations, which is generally computationally expensive due to the broad parameter surveys and the complicated couplings among astrophysical and relativistic effects. Recently, a simple model of an equatorial ring of magnetized fluid orbiting has been developed to investigate the polarized images of synchrotron emission around Schwarzschild black hole \cite{PZ1} and Kerr black hole \cite{PZ2}. Although only the emission from a single radius $r_s$ is considered, this model can clearly reveal dependence of the polarization signatures on magnetic field configuration, black hole spin and observer inclination. Moreover, with this model, the image of a finite thin disk can be produced by simply summing contributions from individual radii [26, 27]. Thus, this
model has been recently applied to study the polarized image of an equatorial emitting ring in various spacetimes, such as, 4D Gauss-Bonnet black hole \cite{PZ3}, regular black holes \cite{PZ4}, Schwarzschilld-Melvin black hole \cite{PZ7}, and so on.

The observations of galaxies \cite{obs1,obs2} reveal a discrepancy between the observed dynamics and the amount of luminous matter. Theoretically, this discrepancy can be explained by introducing exotic dark matter. However, to date, there is no exact evidence to confirm the existence of dark matter.
Another possible resolution to the discrepancy is a modification of gravity theory. STVG \cite{MOG1} is a kind of  fully covariant modified gravity
(MOG) theories. This MOG theory contains three gravitational fields: the Einstein metric related to massless tensor graviton, a massless scalar graviton and  a massive vector graviton. The STVG-MOG theory \cite{MOG1} has successfully explained the rotation curves of galaxies \cite{MOG4,MOG41s} and the dynamics of galactic clusters \cite{MOG2,MOG3,MOG5}. Moreover, the gravitational waves in this STVG-MOG theory has been studied \cite{MOG6,MOG7}. The Schwarzschild-like  and  Kerr-like black hole solutions in STVG-MOG theory are obtained in \cite{MOG8}. These black hole solutions have an additional MOG parameter yielded a variable gravitational constant. It is of interest to study the observational effects of black holes in STVG-MOG theory  because they could help understand this modified gravity theory and some fundamental issues in physics. Effects of the MOG parameter on the quasinormal modes \cite{MOG17}, black hole shadow \cite{MOG9,MOG19}, black hole thermodynamics \cite{MOG10} and other physical processes \cite{MOG11,MOG12,MOG13,MOG14,MOG15,MOG16,MOG18} have been studied in the spacetime of Schwarzschild-MOG and Kerr-MOG black holes. This paper aims to study the polarization information in the image of a synchrotron emitting ring around a
Kerr-MOG black hole \cite{MOG8}  and to probe the effects of the MOG parameter on the polarization image.

The paper is organized as follows: Section II briefly introduces Kerr-MOG black hole and presents formulas to calculate the observed polarization vector in the image plane of an emitting ring in this spacetime. Section III presents the polarization images of a synchrotron emitting
ring and probes the effects of the MOG parameter on the polarization image. Finally, this paper ends with a summary.

\section{Observed polarization field in a Kerr-MOG black hole spacetime}
The STVG theory is a covariant modified theory of gravity and its action is composed of scalar, tensor and vector fields \cite{MOG1}
\begin{eqnarray}\label{action0}
S=S_G+S_{\phi}+S_S,
\end{eqnarray}
with
\begin{eqnarray}
S_G&=&-\frac{1}{16\pi}\int \sqrt{-g}\frac{R}{G}d^4x,\nonumber\\
S_{\phi}&=&-\frac{1}{4\pi}\int \sqrt{-g}\bigg[\frac{1}{4}B_{\mu\nu}B^{\mu\nu}+V(\phi_{\mu}\phi^{\mu})\bigg]d^4x,\nonumber\\
S_{S}&=&-\int \sqrt{-g}\frac{1}{G}\bigg[\frac{1}{2}g^{\alpha\beta}\bigg(\frac{\nabla_{\alpha}G\nabla_{\beta}G}{G^2}
+\frac{\nabla_{\alpha}\mu\nabla_{\beta}\mu}{\mu^2}\bigg)+\frac{V(G)}{G^2}+\frac{V(\mu)}{\mu^2}\bigg]d^4x.
\end{eqnarray}
$S_G$ corresponds to the Einstein gravity action and $R$ is Ricci scalar. $S_{\phi}$ is the action of the massive vector field $\phi_{\mu}$ and $B_{\mu\nu}=\partial_{\mu}\phi_{\nu}-\partial_{\nu}\phi_{\mu}$. $G$ is a scalar
field $G=G_N(1+\alpha)$ corresponding to a spin 0 massless graviton, $G_N$ is Newton's gravitational
constant and $\alpha$ is a dimensionless parameter. In this modified theory, the mass of the vector field $\phi_{\mu}$ is also an effective spin 0
scalar field $\mu$. $V (G)$ and $V(\mu)$ are self-interaction potentials of $\mu$
and $G$ fields, respectively.
The spacetime of a rotating black hole in the STVG-MOG theory \cite{MOG1} can be described by the so-called
 Kerr-MOG metric with the form in Boyer-Lindquist coordinate \cite{MOG8}
\begin{eqnarray}\label{Metric01}
ds^2=-\frac{\Delta\rho^2}{\Xi}dt^2+\frac{\rho^2}{\Delta}dr^2+\Sigma{d\theta^2}+\frac{\Xi\sin\theta^2}{\rho^2}(d\phi-\omega{dt})^2,
\end{eqnarray}
where
\begin{eqnarray}\label{Metric02}
&&\Delta=r^2-2GMr+a^2+\alpha{G_NGM^2},\quad\quad \rho^2=r^2+a^2\cos\theta^2,  \\ \nonumber
&&\omega=\frac{a(a^2+r^2-\Delta)}{\Xi}, \quad\quad\quad\quad\quad\quad\quad\quad \Xi=\left(r^2+a^2\right)^2-\Delta{a^2}\sin\theta^2.
\end{eqnarray}
Here $M$ and $a$ are the mass parameter and the spin parameter of Kerr-MOG black hole, respectively. The Kerr-MOG metric can reduce to the Kerr metric in the limit $\alpha=0$. Thus, the parameter $\alpha$ can be used to measure the deviation of Kerr-MOG black hole in STVG-MOG theory from Kerr black hole in general relativity. The Arnowitt-Deser-Misner (ADM) mass and the angular momentum of the Kerr-MOG black hole are given by \cite{MOG15}
\begin{eqnarray}\label{ADM}
\mathcal{M}=(1+\alpha)M, \quad\quad J=\mathcal{M}a.
\end{eqnarray}
With the ADM mass $\mathcal{M}$, the function $\Delta$ can be rewritten as
\begin{eqnarray}\label{Metric03}
\Delta=r^2-2G_N\mathcal{M}r+a^2+\frac{\alpha}{1+\alpha}G^2_N\mathcal{M}^2.
\end{eqnarray}
The outer and inner horizon radiuses are the roots of $\Delta=0$, i.e.,
\begin{eqnarray}\label{horizon}
r_\pm=\mathcal{M}\pm\sqrt{\frac{\mathcal{M}^2}{1+\alpha}-a^2}.
\end{eqnarray}
Here we set $G_N=1$ without loss of generality. The extremal limit for Kerr-MOG black hole is $\mathcal{M}^2=(1+\alpha)a^2$. Similarly, the Kerr-MOG metric is singular at $\rho^2=0$ as in the Kerr case.

Now, the polarization vectors are to be studied for photons emitted from the ring with radius $r_s$ around a Kerr-MOG black
hole. A synchrotron emitting ring is assumed to lie in the equatorial plane $(\theta_s=\frac{\pi}{2})$. In the
local ZAMO frame of the point $P$ in the ring, the four-vector components $V^{(a)}$ of emitter are related to the vector $V_{\mu}$ in Kerr-MOG spacetime by \cite{PZ1,PZ2}
\begin{eqnarray}\label{trans01}
V^{(a)}=\eta^{(a)(b)}e^\mu_{(b)}V_\mu,
\end{eqnarray}
where $\eta^{(a)(b)}$ is the flat Minkowski metric and $e^\mu_{(b)}$ is the zero-angular-momentum-observer (ZAMO) tetrad
\begin{equation}\label{tetrad}
e^\mu_{(b)}=\left[
\begin{array}{cccc}
\frac{1}{r_s}\sqrt{\frac{\Xi_s}{\Delta_s}} & 0 & \frac{\omega_s}{r_s}\sqrt{\frac{\Xi_s}{\Delta_s}} & 0 \\
0 & \frac{\sqrt{\Delta_s}}{r_s} & 0 & 0 \\
0 & 0 & \frac{r_s}{\sqrt{\Xi_s}} & 0 \\
0 & 0 & 0 & -\frac{1}{r_s}
\end{array}
\right].
\end{equation}
For a boosted emitter in the local orthonormal ZAMO frame, its boosting velocity is assumed to be in the $r-\phi$ plane and has a form
\begin{eqnarray}\label{boost01}
\vec{\beta}=\beta_\nu[\cos\chi(\hat{r})+\sin\chi(\hat{\phi})].
\end{eqnarray}
The vectors of point emitter in the the boosted
orthonormal frame can be obtained by a Lorentz transformation $\Lambda^{(a)}_{\;\;\;(b)}$ from
the ZAMO frame, i.e.,
\begin{eqnarray}\label{trans02}
V^{'(a)}=\Lambda^{(a)}_{\;\;\;(b)}V^{(b)},
\end{eqnarray}
where
\begin{equation}\label{boost02}
\Lambda^{(a)}_{\;\;\;(b)}=\left[
\begin{array}{cccc}
\gamma & -\beta_\nu\gamma\cos\chi & -\beta_\nu\gamma\sin\chi & 0 \\
-\beta_\nu\gamma\cos\chi & (\gamma-1)\cos\chi^2+1 & (\gamma-1)\sin\chi\cos\chi & 0 \\
-\beta_\nu\gamma\sin\chi & (\gamma-1)\sin\chi\cos\chi & (\gamma-1)\sin\chi^2+1 & 0 \\
0 & 0 & 0 & 1
\end{array}
\right],
\end{equation}
and $\gamma$  is the Lorentz factor.
Thus, the vector of the point source in black hole spacetime can be obtained from the boosted orthonomal frame by an inverse transformation
\begin{eqnarray}\label{trans03}
V^\mu=e_{(c)}^\mu\Lambda^{\;\;\;(c)}_{(a)}V^{'a}.
\end{eqnarray}
where $\Lambda^{\;\;\;(b)}_{(a)}$ is the inverse matrix of $\Lambda^{(a)}_{\;\;\;(b)}$. Here, the orientation $(\hat{x},\hat{y},\hat{z})$ is equivalent to $(\hat{r},\hat{\phi},\hat{\theta})$.

To study the polarized image of an equatorial emitting ring around a Kerr-MOG black hole, the null geodesic describing photon propagation must be solved firstly. Using the Hamilton-Jacobi approach, the null geodesic equation in Kerr-MOG black hole spacetime can be expressed as
\begin{eqnarray}\label{geodesic}
&&\frac{\rho^2}{E}p^t=\frac{r^2+a^2}{\Delta}(r^2+a^2-a\lambda)+a(\lambda-a\sin\theta^2),  \nonumber \\
&&\frac{\rho^2}{E}p^\phi=\frac{a}{\Delta}(r^2+a^2-a\lambda)+\frac{\lambda}{\sin\theta^2}-a, \nonumber \\
&&\frac{\rho^2}{E}p^r=\pm_r\sqrt{\mathcal{R}(r)},  \nonumber \\
&&\frac{\rho^2}{E}p^\theta=\pm_\theta\sqrt{\Theta(\theta)}.
\end{eqnarray}
The conserved quantities $\lambda$ and $\eta$ correspond to the energy-rescaled angular momentum parallel to the axis of symmetry and Carter constant, respectively.  The radial  potential $\mathcal{R}(r)$ and the angular potential $\Theta(\theta)$ can be expressed as
\begin{eqnarray}\label{potential}
&&\mathcal{R}(r)=\left(r^2+a^2-a\lambda\right)^2-\Delta\left[\eta+(a-\lambda)^2\right],   \nonumber \\
&&\Theta(\theta)=\eta+a^2\cos\theta^2-\lambda^2\cot\theta^2.
\end{eqnarray}
The photon trajectory is determined by its initial
position, the conserved quantities $\lambda,\eta$, and the signs $\pm_r$, $\pm_\theta$ of
its initial motion. The photon's four-momentum can be given by
\begin{eqnarray}\label{momentum01}
p_\mu{dx^\mu}=-dt\pm_r\frac{\sqrt{\mathcal{R}(r)}}{\Delta(r)}dr\pm_\theta\sqrt{\Theta(\theta)}d\theta+\lambda{d\phi}.
\end{eqnarray}
In the Kerr-MOG black hole spacetime \eqref{Metric01}, the celestial coordinates $(x,y)$ for the photon's arrival position on the observer's screen
are
\begin{eqnarray}\label{xy01}
x=-\frac{\lambda}{\sin\theta_o}, \quad\quad\quad  y=\pm_o\sqrt{\Theta(\theta)},
\end{eqnarray}
where angel $\theta_o$ is the observer's polar inclination from the normal direction of the accretion disk and $\pm_o$ is the sign of $p^\theta$.

As the photon emitted from the initial position $(r_s,\theta_s)$ moves along the null geodesic to the observer at position $(r_o,\theta_o)$, its trajectory is given by the null geodesic equation \eqref{geodesic} \cite{PZ2,Math1}
\begin{eqnarray}\label{integra01}
 I_r\equiv\fint_{r_s}^{r_o}\frac{dr}{\pm_r\sqrt{\mathcal{R}(r)}}=\fint_{\theta_s}^{\theta_o}\frac{d\theta}{\pm_r\sqrt{\Theta(\theta)}}\equiv G_\theta.
\end{eqnarray}
Here, the slash denotes that the sign of $\pm_{r}=sign(p^r)$ or $\pm_\theta=sign(p^\theta)$ in the path integral
change at radial or angular turning point.
For a photon's trajectory with $m$ turning points in $\theta$ and $\theta_s =\frac{\pi}{2}$, the null geodesic equation can be simplified as \cite{PZ2,Math1}
\begin{eqnarray}\label{xy02}
\sqrt{-u_{-}a^2}I_r+sign(\beta)F_o=2mK\left(\frac{u_+}{u_-}\right),
\end{eqnarray}
with
\begin{eqnarray}\label{xy03}
F_o=F\left(\arcsin\frac{\cos\theta_o}{\sqrt{u_+}}\Big|\frac{u_+}{u_-}\right),  \quad\quad u_\pm=\Delta_\theta\pm\sqrt{\Delta_\theta^2+\frac{\eta}{a^2}},  \quad\quad  \Delta_\theta=\frac{1}{2}(1-\frac{\eta+\lambda^2}{a^2}),
\end{eqnarray}
where $F$ and $K$ denote the first-kind incomplete and complete elliptic integrals, respectively. Actually, from the geodesic equation (\ref{xy02}), an inversion formula for the emission radius $r_s(I_r)$ can be obtained \cite{Math1}
\begin{eqnarray}\label{xy04}
r_s(I_r)=\frac{r_4r_{31}-r_3r_{41}sn^2(\frac{1}{2}\sqrt{r_{31}r_{42}}I_r-\mathcal{F}_o\big|k)}{r_{31}-r_{41}sn^2(\frac{1}{2}\sqrt{r_{31}r_{42}}I_r-\mathcal{F}_o\big|k)},
\end{eqnarray}
where
\begin{eqnarray}\label{xy05}
\mathcal{F}_o=F\left(\arcsin\sqrt{\frac{r_{31}}{r_{41}}}\Big|k\right),  \quad\quad\quad\quad   k=\frac{r_{32}r_{41}}{r_{31}r_{42}},   \quad\quad\quad\quad  r_{ij}=r_j-r_j.
\end{eqnarray}
Here $ r_1,r_2,r_3,r_4 $ are the four roots of radial potential $\mathcal{R}(r)$ and $sn(q|k) $ is Jacobi elliptic function. In the Kerr-MOG black hole spacetime (\ref{Metric01}), the four roots $r_1,r_2,r_3,r_4$ can also expressed as
\begin{eqnarray}\label{xy08}
&&r_{1,2}=-z\mp\sqrt{-\frac{\mathcal{A}}{2}-z^2+\frac{\mathcal{B}}{4z}},  \quad\quad\quad\quad  r_{3,4}=z\mp\sqrt{-\frac{\mathcal{A}}{2}-z^2+\frac{\mathcal{B}}{4z}},
\end{eqnarray}
where
\begin{eqnarray}\label{xy07}
&&z=\sqrt{\frac{\omega_++\omega_-}{2}-\frac{\mathcal{A}}{6}}>0,   \quad\quad\quad    \omega_\pm=\sqrt[3]{-\frac{\mathcal{Q}}{2}\pm\sqrt{\left(\frac{\mathcal{P}}3{}\right)^3+\left(\frac{\mathcal{Q}}{2}\right)}},\\ \nonumber
&&\mathcal{P}=-\frac{\mathcal{A}^2}{12}-\mathcal{C},  \quad\quad\quad\quad\quad\quad\quad\quad   \mathcal{Q}=-\frac{\mathcal{A}}{3}\left[\left(\frac{\mathcal{A}}{6}\right)^2-\mathcal{C}\right]-\frac{\mathcal{B}^2}{8},
\end{eqnarray}
with
\begin{eqnarray}\label{xy06}
&&\mathcal{A}=a^2-\eta-\lambda^2, \\ \nonumber
&&\mathcal{B}=\frac{2\mathcal{M}}{1+\alpha}\{\eta+(\lambda-a)^2+\alpha[\eta+(\lambda-a)^2]\}, \\ \nonumber
&&\mathcal{C}=-a^2\eta-\frac{\alpha\mathcal{M}^2}{1+\alpha}[\eta+(\lambda-a)^2].
\end{eqnarray}
Combining \eqref{xy02} and \eqref{xy04}, one can numerically compute the set of celestial coordinates $(x,y)$ for a photon emitted on an equatorial ring with a radius $r_s$.  The photon four-momentum at the source $(r_s,\theta_s=\frac{\pi}{2})$ can be expressed as
\begin{eqnarray}\label{momentum02}
p_t=-1,  \quad\quad\quad\quad  p_r=\pm_r\frac{\sqrt{\mathcal{R}(r_s)}}{\Delta_s},  \quad\quad\quad\quad  p_\theta=\pm_s\sqrt{\eta},   \quad\quad\quad\quad  p_\phi=\lambda,
\end{eqnarray}
where the sign of $p^{\theta}$ at the source $\pm_s=(-1)^m\pm_o$. The sign of $\pm_r$ can be computed by a semi-analytic calculation \cite{Math1}. With the determined sign of $p^r$,  the four-momentum $p^{\mu}$ of photon at the source can be calculated by
\begin{eqnarray}\label{momentum03}
&&p^t=\frac{1}{r_s^2}\left[\frac{r_s^2+a^2}{\Delta_s}(r_s^2+a^2-a\lambda)+a(\lambda-a)\right], \quad\quad  p^r=\pm_r\frac{1}{r_s^2}\sqrt{\mathcal{R}(r_s)},  \\ \nonumber
&&p^\phi=\frac{1}{r_s^2}\left[\frac{a}{\Delta_s}(r_s^2+a^2-a\lambda)+\lambda-a\right],   \quad\quad\quad\quad\quad p^\theta=\pm_s\frac{\sqrt{\eta}}{r_s^2}.
\end{eqnarray}

Converting the four-momentum $p^\mu$ to the local frame of the emitter  by \eqref{trans02}, the local photon polarization at the source can be obtained. In the local frame, one has $f^{t}=0$. For the synchrotron radiation,  the spatial components of the photon polarization vector at the source $\vec{f}=(f^{r},f^{\theta},f^{\phi})$ are related to the local three-momentum $\vec{p}=(p^{r},p^{\theta},p^{\phi})$ and the local
magnetic field $\vec{B}=(B^{r},B^{\theta},B^{\phi})$ by
\begin{eqnarray}\label{polarized01}
\vec{f}=\frac{\vec{p}\times\vec{B}}{|\vec{p}|}.
\end{eqnarray}
 The photon polarization vector at the source $f^\mu$ in Boyer-Linquist coordinates can be computed through the transformation \eqref{trans03}.
 With the angle $\zeta$ between $\vec{p}$ the magnetic field $\vec{B}$
\begin{eqnarray}\label{polarized03}
\sin\zeta=\frac{|\vec{p}\times\vec{B}|}{|\vec{p}||\vec{B}|},
\end{eqnarray}
the normalized polarization vector satisfies
\begin{eqnarray}\label{polarized04}
f^\mu{f_\mu}=\sin\zeta^2|\vec{B}|^2.
\end{eqnarray}
In the propagation of photon in the Kerr-MOG black hole spacetime (\ref{Metric01}), the polarization vector $f^{\mu}$ obeys
\begin{eqnarray}\label{polarized02}
f^\mu{p_\mu}=0, \quad\quad\quad  p^\mu\nabla_\mu{f^\nu}=0.
\end{eqnarray}
According to Walker-Penrose theorem \cite{PWconstant,Chandrasekhar}, along  null geodesic in the Kerr-MOG geometry (\ref{Metric01}), there is a conserved complex quantity
\begin{eqnarray}\label{PW constant01}
\kappa=p^if^j(l_in_j-l_jn_i-m_i\bar{m}_j+\bar{m}_im_j)\Psi_2^{\left(-\frac{1}{3}\right)},
\end{eqnarray}
with
\begin{eqnarray}\label{PW constant02}
&&\kappa=\kappa_1+i \kappa_2=(A-i{B})\Psi_2^{-\frac{1}{3}}, \\ \nonumber
&&A=(p^tf^r-p^rf^t)+a\sin\theta^2(p^rf^\phi-p^\phi{f^r}), \\ \nonumber
&&B=\left[(r^2+a^2)(p^\phi{f^\theta}-p^\theta{f^\phi})-a(p^tf^\theta-p^\theta{f^t})\right]\sin\theta.
\end{eqnarray}
$\Psi_2$ is Weyl scalar and has a form
\begin{eqnarray}\label{PW constant03}
\Psi_2=-\frac{\mathcal{M}}{(r-i{a}\cos\theta)^3}\bigg[1-\frac{\alpha\mathcal{M}}{(1+\alpha)(r+i{a}\cos\theta)}\bigg].
\end{eqnarray}
Making use of the celestial coordinates $(x,y)$ and the Walker-Penrose constant $\kappa$ at the source $(r_s,\theta_s=\frac{\pi}{2})$,  the  polarization vector on the observer's screen can be obtained by \cite{Chandrasekhar,Himwich}
\begin{eqnarray}\label{polarized05}
f^x=\frac{y\kappa_2-\mu\kappa_1}{\mu^2+y^2}, \quad\quad\quad f^y=\frac{y\kappa_1+\mu\kappa_2}{\mu^2+y^2},
\quad\quad\quad \mu=-(x+a\sin\theta_o).
\end{eqnarray}
\begin{figure}
\centering
\includegraphics[width=14cm]{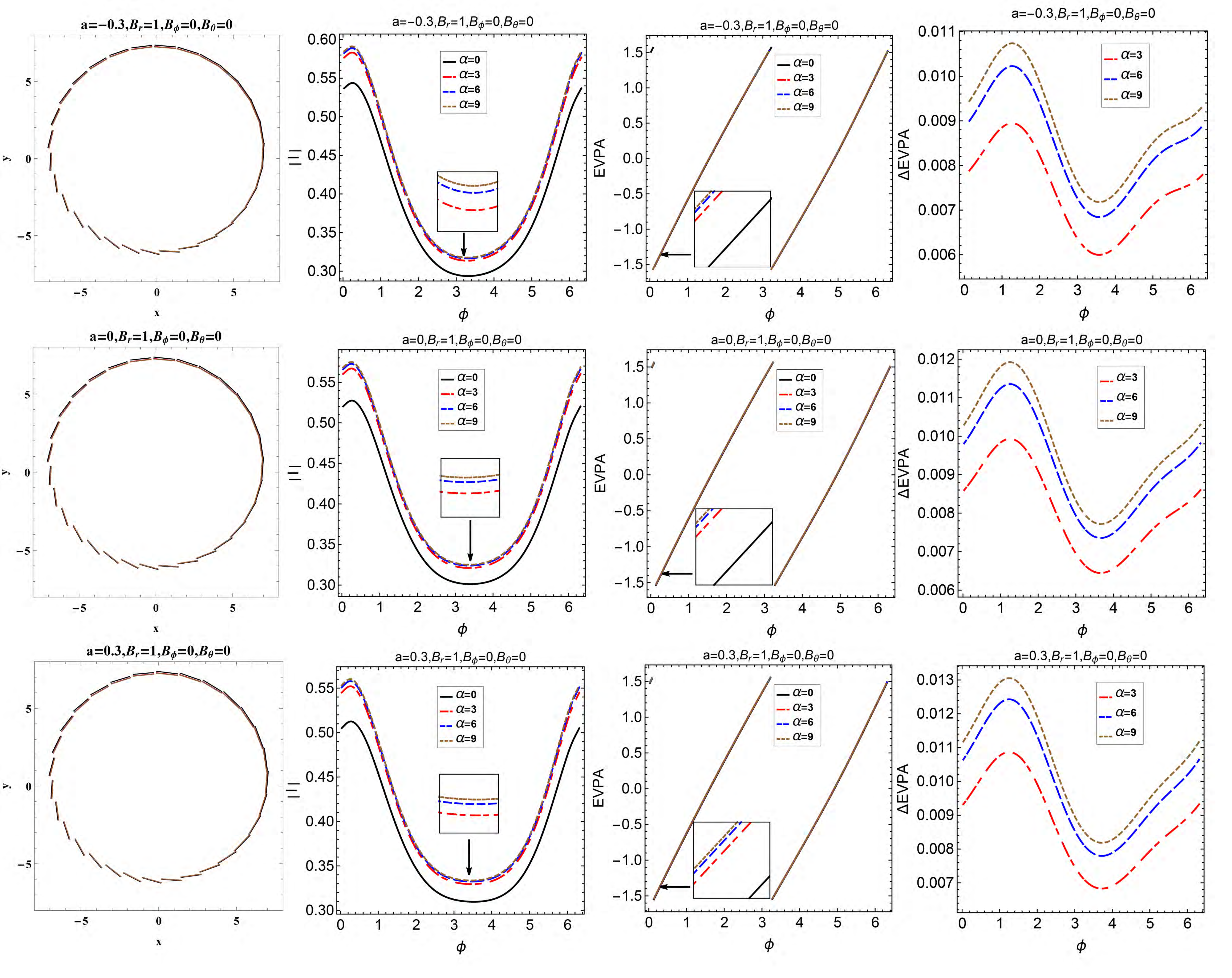}
\caption{Effects of the MOG parameter $\alpha$ on the polarized vector and EVPA in the Kerr-MOG black hole spacetime \eqref{Metric01} for the magnetic field owned only the radial component $B_r$. Here we set $\mathcal{M}=1$, $r_s=6$, $\theta_o=20^{\circ}$, $\beta_\nu=0.3$, and $\chi=-90^{\circ}$.  }
\label{f1}
\end{figure}
\begin{figure}
\centering
\includegraphics[width=14cm]{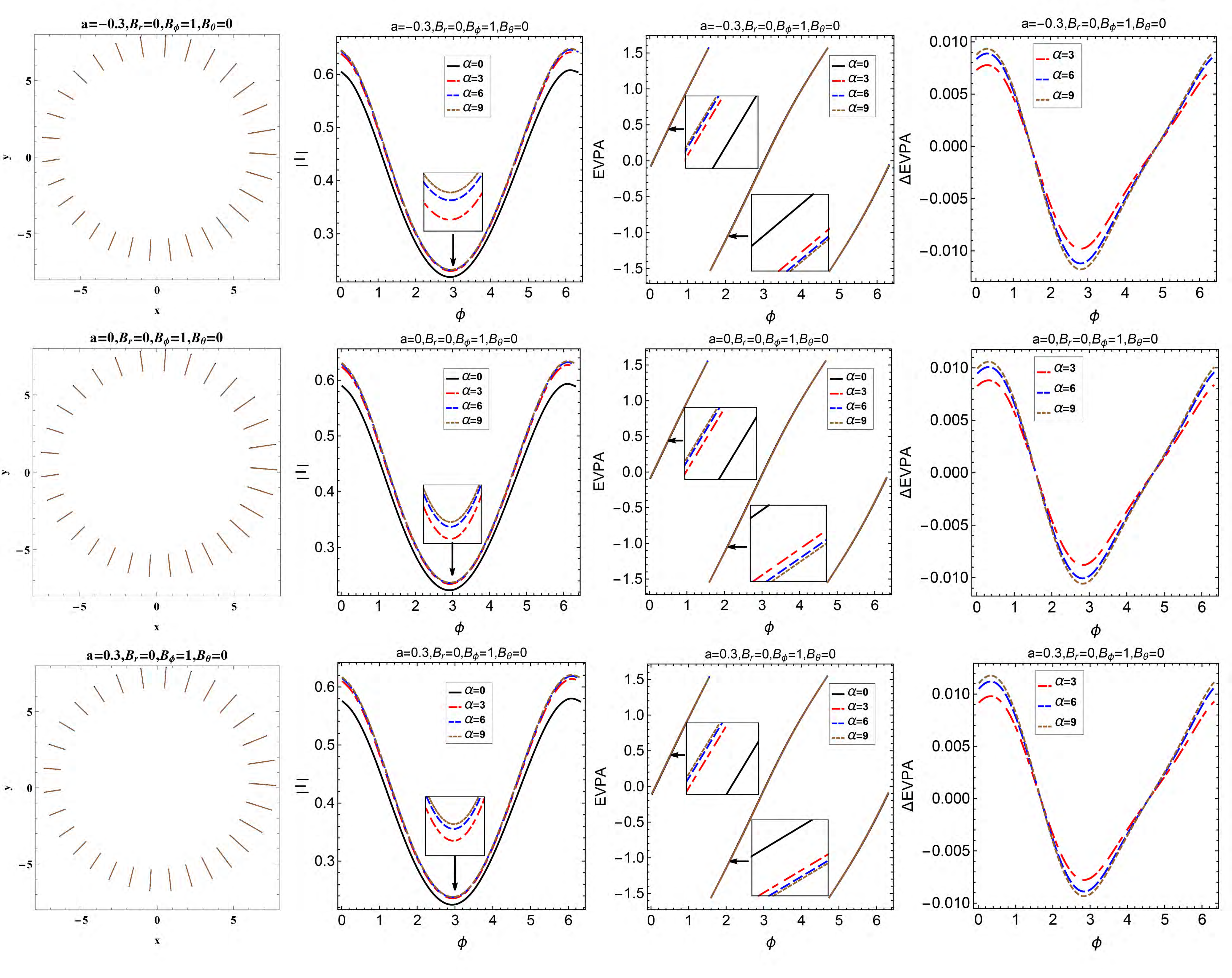}
\caption{Effects of the MOG parameter $\alpha$ on the polarized vector and EVPA in the Kerr-MOG black hole spacetime \eqref{Metric01} for the equatorial magnetic field owned only the angular component $B_{\phi}$. Here we set $\mathcal{M}=1$, $r_s=6$, $\theta_o=20^{\circ}$, $\beta_\nu=0.3$, and $\chi=-90^{\circ}$.}
\label{f2}
\end{figure}
In general, the intensity of linear polarization for synchrotron radiation that reaches the observer from the source position can be approximated as \cite{PZ1,PZ2}
\begin{eqnarray}\label{intens01}
|I|=g^{3+\alpha_\nu}l_p|\vec{B}|^{1+\alpha_\nu}\sin\zeta^{1+\alpha_\nu},
\end{eqnarray}
where $g$ is the redshift factor measured by the ratio of the photon energies at the observer $E_o=1$ and at the emitter $E_s=p^{(t)}$.
The power $\alpha_\nu$ depends on the properties of the accretion disk. Here, we set $\alpha_\nu=1$ as in refs.\cite{PZ1,PZ2}. The quantity $l_p=\frac{p_s^{(t)}}{p_s^{(z)}}H$ is the geodesic path length through the emitting material. $H$ is the height of the disk which can be taken to be a constant for simplicity. Thus, the observed components of photon polarization vector are
\begin{eqnarray}\label{intens02}
f_{obs}^x=\sqrt{l_p}g^2|B|\sin\zeta f^x,  \quad\quad\quad f_{obs}^y=\sqrt{l_p}g^2|B|\sin\zeta f^y.
\end{eqnarray}
Finally, the total polarization intensity and the electric vector position angle (EVPA) can be given by
\begin{eqnarray}\label{intens03}
I=\left(f_{obs}^x\right)^2+\left(f_{obs}^y\right)^2,  \quad\quad\quad \rm{EVPA}=\frac{1}{2}\arctan\frac{U}{Q},
\end{eqnarray}
where $Q$ and $U$ are the Stokes parameters
\begin{eqnarray}\label{intens04}
Q=\left(f_{obs}^y\right)^2-\left(f_{obs}^x\right)^2,  \quad\quad\quad U=-2f_{obs}^xf_{obs}^y.
\end{eqnarray}
For the Kerr-MOG black hole spacetime (\ref{Metric01}), combining photon geodesic
with Eqs. \eqref{PW constant01}, \eqref{intens01}, \eqref{intens02}, \eqref{intens03} and \eqref{intens04}, the polarization intensity and EVPA in the pixel related to the point source can be obtained. Repeating similar operations along the emitting ring, the total polarization image of the emitting ring around a Kerr-MOG black hole black hole and the corresponding the effects of MOG parameter can be presented.

\section{Effects of MOG parameter on the polarized image of an equatorial emitting ring around a Kerr-MOG black hole}

\begin{figure}[htb!]
\includegraphics[width=14cm]{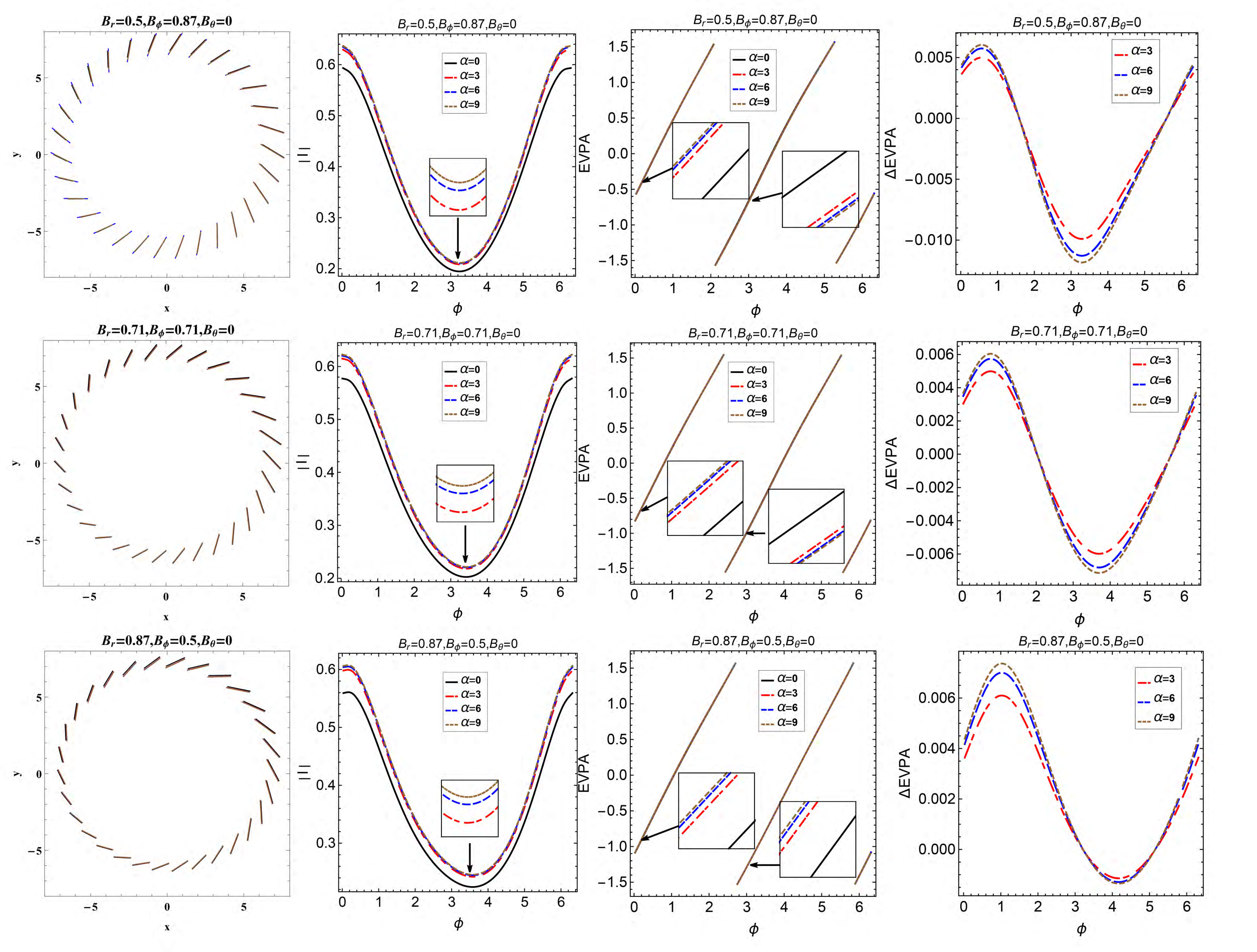}
\caption{Effects of the MOG parameter $\alpha$ on the polarized vector and EVPA in the Kerr-MOG black hole spacetime \eqref{Metric01} for the equatorial magnetic field. Here $r_s=6$, $a=-0.3$, $\theta_o=20^{\circ}$, $\beta_\nu=0.3$, and $\chi=-90^{\circ}$.}
\label{f3}
\end{figure}

\begin{figure}[htb!]
\includegraphics[width=14cm]{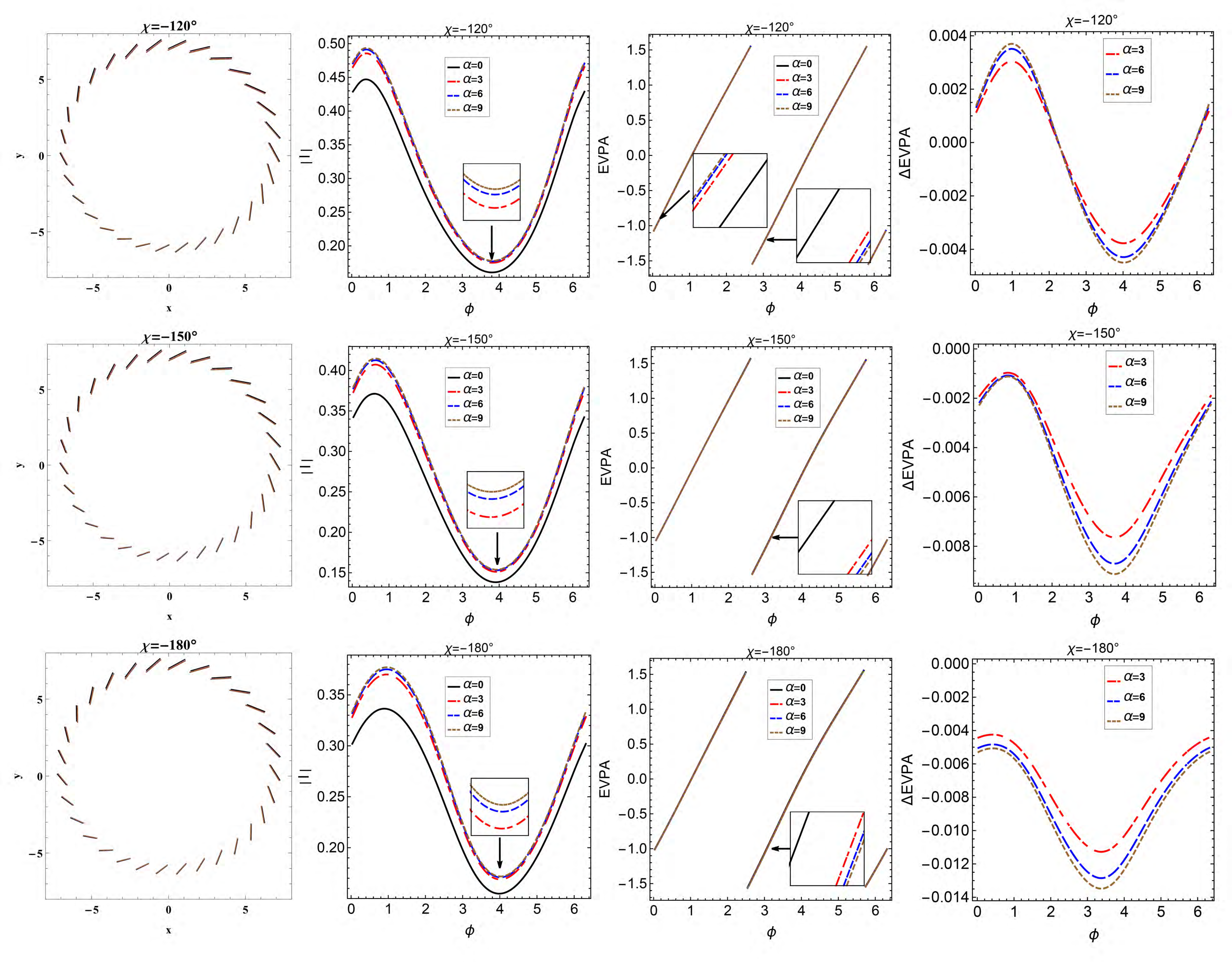}
\caption{Effects of the MOG parameter $\alpha$ on the polarized vector and EVPA in the Kerr-MOG black hole spacetime \eqref{Metric01} for the different  fluid direction angle $\chi$. Here $r_s=6$, $a=-0.3$, $\theta_o=20^{\circ}$, $\beta_\nu=0.3$, $B_r=0.87$, $B_\phi=0.5$ and $B_\theta=0$.}
\label{f4}
\end{figure}

\begin{figure}[htb!]
\includegraphics[width=14cm]{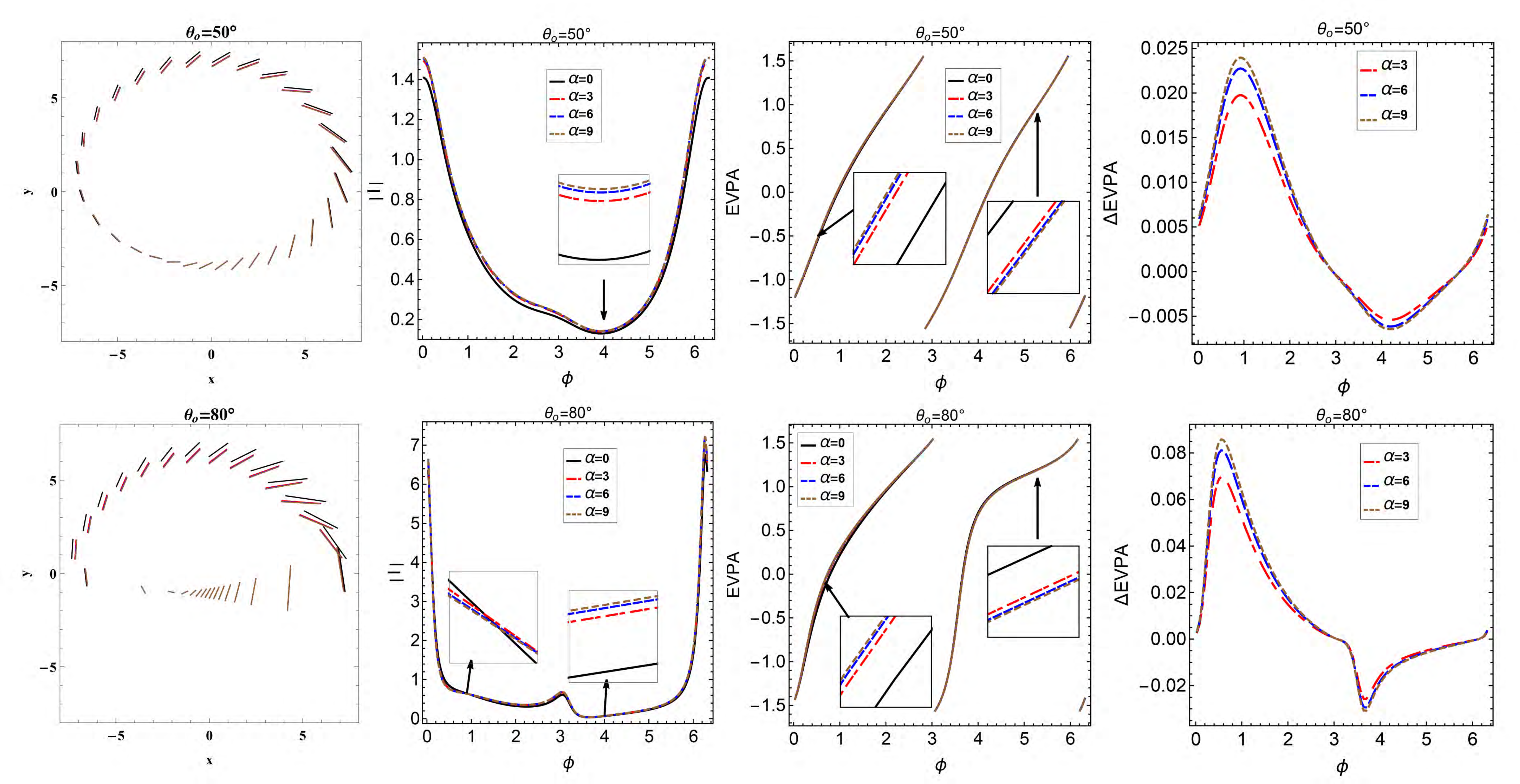}
\caption{Effects of the MOG parameter $\alpha$ on the polarized vector and EVPA  in the Kerr-MOG black hole spacetime \eqref{Metric01} for the different observer inclination angle $\theta_o$. Here $r_s=6$, $a=-0.3$, $\beta_\nu=0.3$, $\chi=-90^{\circ}$, $B_r=0.87$, $B_\phi=0.5$ and $B_\theta=0$.}
\label{f5}
\end{figure}

\begin{figure}[htb!]
\includegraphics[width=14cm]{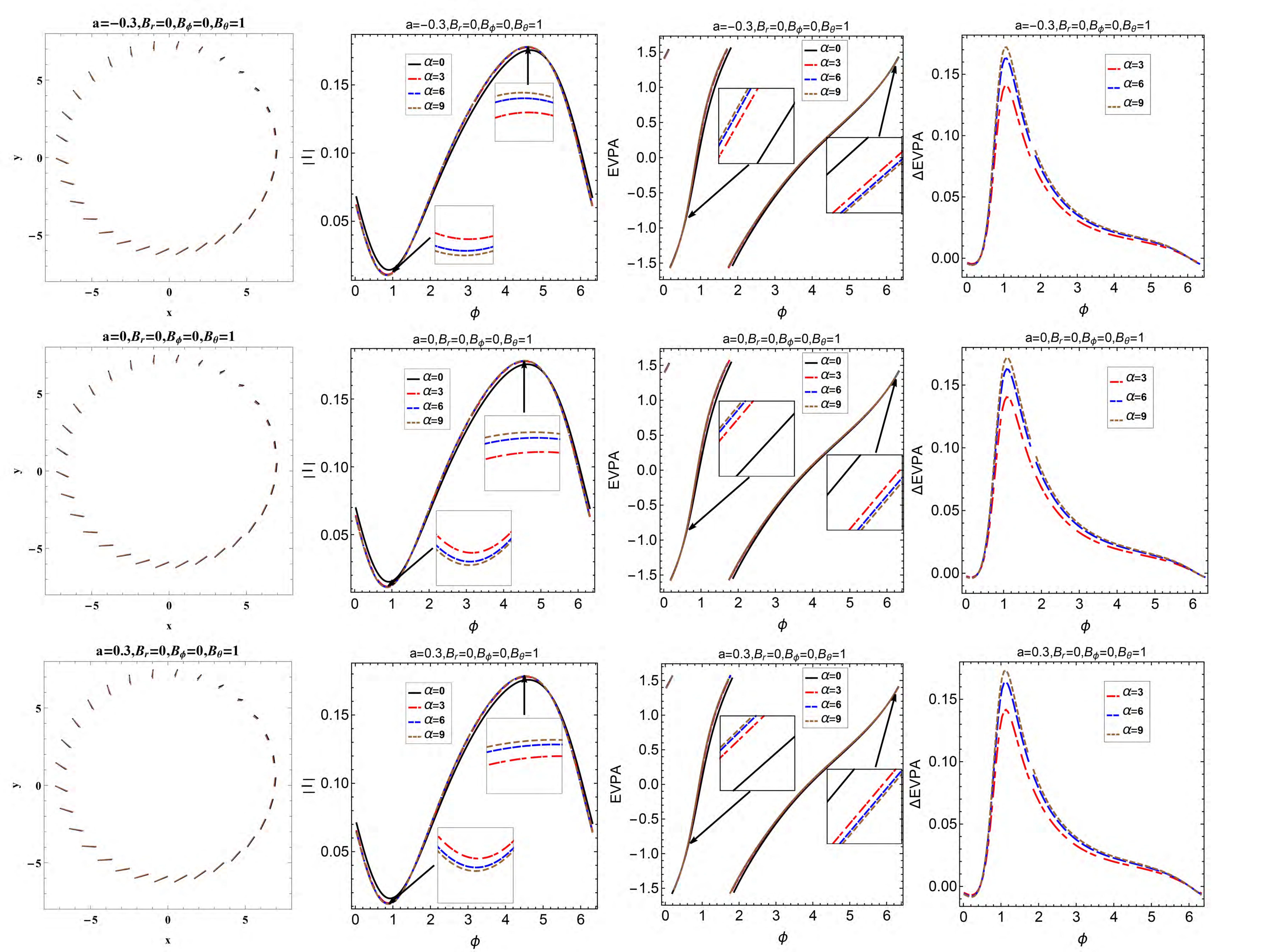}
\caption{Effects of the MOG parameter $\alpha$ on the polarized vector and EVPA in the Kerr-MOG black hole spacetime \eqref{Metric01} for different $a$ in the case with the magnetic field owned only the vertical component $B_\theta$. Here $r_s=6$, $\theta_o=20^{\circ}$, $\chi=-90^{\circ}$, $\beta_\nu=0.3$,  and $B_\theta=1$.}
\label{f6}
\end{figure}

\begin{figure}[htb!]
\includegraphics[width=14cm]{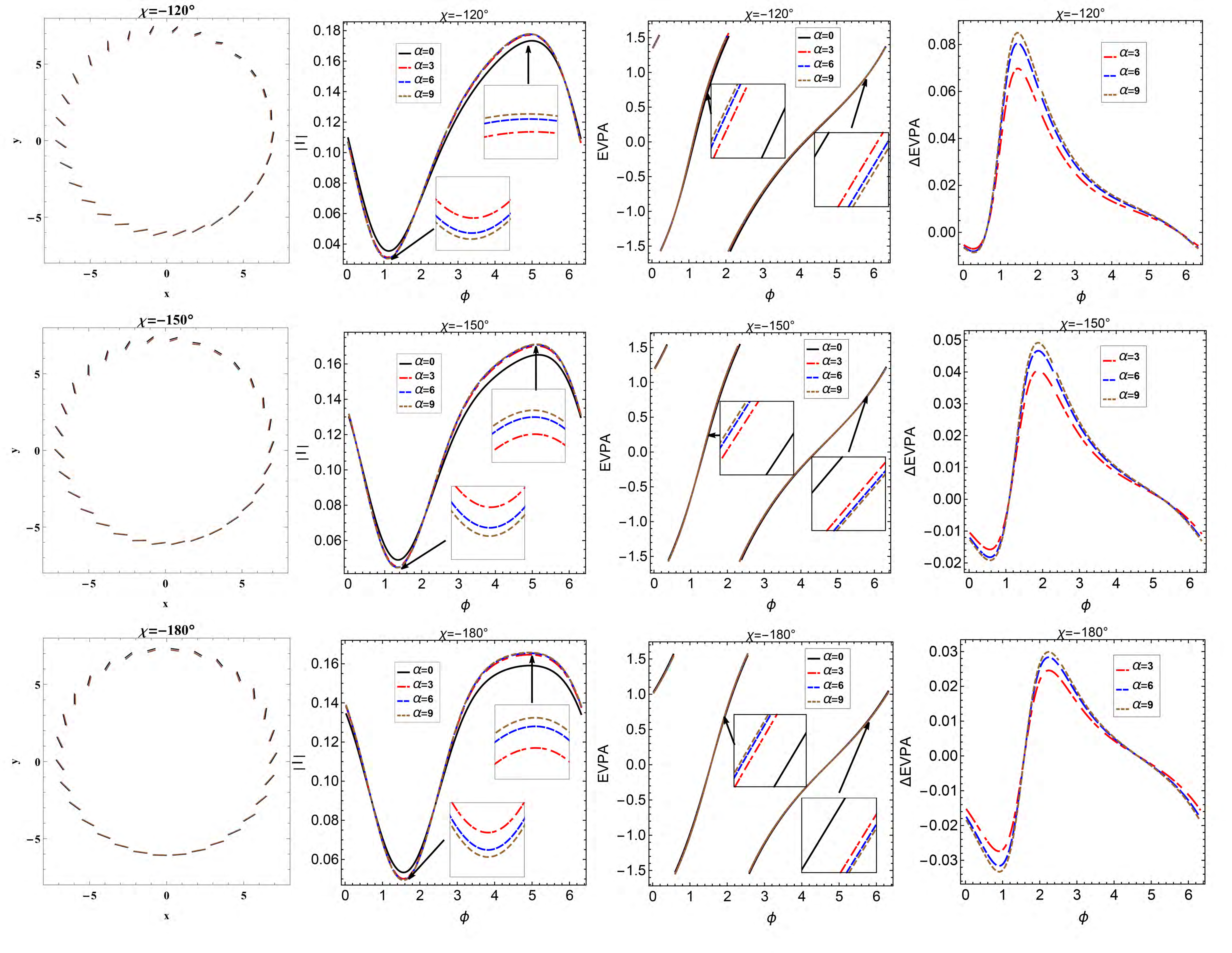}
\caption{Effects of the MOG parameter $\alpha$ on the polarized vector and EVPA in the Kerr-MOG black hole spacetime \eqref{Metric01} for different $\chi$ in the case where magnetic field has only the vertical component $B_\theta$. Here $r_s=6$, $a=-0.3$, $\theta_o=20^{\circ}$, $\beta_\nu=0.3$, and $B_\theta=1$.}
\label{f7}
\end{figure}

\begin{figure}[htb!]
\includegraphics[width=14cm]{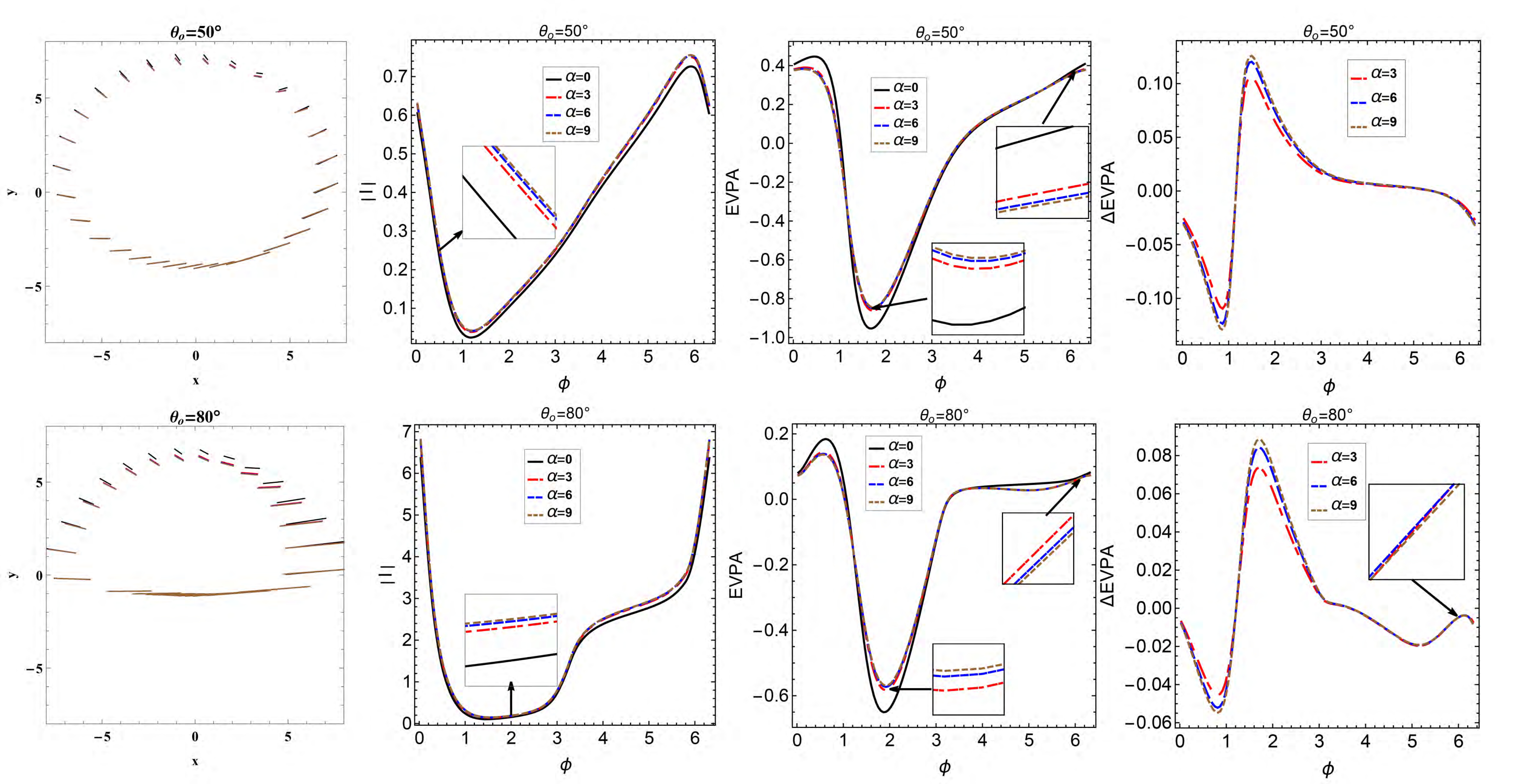}
\caption{Effects of the MOG parameter $\alpha$ on the polarized vector and EVPA in the Kerr-MOG black hole spacetime \eqref{Metric01} for different $\theta_o$ in the case where magnetic field has only the vertical component $B_\theta$. Here $r_s=6$, $a=-0.3$, $\beta_\nu=0.3$, $\chi=-90^{\circ}$, and $B_\theta=1$.}
\label{f8}
\end{figure}

Figs.(\ref{f1})-(\ref{f8}) present the polarization vector distribution in the image of the emitting ring with radius $r_s=6$ around a Kerr-MOG black hole. Results show that the polarization vector distribution in the image depends on not only the magnetic field configuration, the motion of fluid particle and the observer's inclination angle, but also on the spin parameter and the MOG parameter $\alpha$ of black hole.
Figs.(\ref{f1})-(\ref{f5}) present the change of the polarized vector and EVPA with the MOG parameter $\alpha$ in the case the magnetic field lies in the equatorial plane for the fixed parameters $r=6$, $\theta_o=20^{\circ}$, $\beta_\nu=0.3$ and $\chi=-90^{\circ}$.
For the case with only radial magnetic field, Fig.(\ref{f1}) shows that the polarization intensity and the EVPA increase with the MOG parameter $\alpha$. The quantity $\Delta{\rm EVPA}\equiv{\rm EVPA-EVPA_{K}}$, described the EVPA difference between in Kerr-MOG and Kerr spacetimes, increases with the parameter $\alpha$. In the cases with different spin parameters, the change of the polarization image feature with $\alpha$  is qualitatively similar in the Kerr-MOG spacetime. For the case with only angular magnetic field, Fig.(\ref{f2}) indicates that the polarization intensity increase with the MOG parameter $\alpha$. However, EVPA increases with $\alpha$ as the azimuthal coordinate $\phi$ lies in the region $(0,\frac{\pi}{2})\cup (\frac{3\pi}{2},2\pi)$, but decreases with $\alpha$ as the azimuthal coordinate $\phi\in(\frac{\pi}{2},\frac{3\pi}{2})$.
For the equatorial magnetic field  with nonzero radial and angular components, from Fig.(\ref{f3}), the polarization intensity still increases with $\alpha$. The change of EVPA becomes more complicated.  With the increasing of the ratio $B_r/B_\phi$, the region where EVPA increases with $\alpha$ becomes broad so that EVPA finally becomes a increasing function of $\alpha$.

Fig.(\ref{f4}) presents the effects of MOG parameter $\alpha$ on the polarized images for different fluid direction angle $\chi$. For the different $\chi$, the polarization intensity still increases with $\alpha$. However, as the angle $\chi$ changes from $-120^{\circ}$ to $\chi=-180^{\circ}$, the region where EVPA increases with $\alpha$ becomes narrow and then EVPA finally becomes a decreasing function of $\alpha$.
The effects of MOG parameter $\alpha$ on the polarized vector and EVPA for different observer inclination angle $\theta_o$ are shown in Fig.(\ref{f5}). It is shown that with the increase of the observer inclination angle $\theta_o$, the region where polarization intensity and EVPA increase with $\alpha$ becomes narrow.

Figs.(\ref{f6})-(\ref{f8}) also present the dependence of the polarization intensity and the EVPA on the MOG parameter $\alpha$ in the case where the magnetic field is perpendicular to the equatorial plane. Results show that the effects of MOG parameter $\alpha$ on the polarized vector and EVPA also depend on the azimuthal coordinate $\phi$ of the  point in the emitting ring, the black hole spin parameter, the fluid direction angle and the observer inclination angle. The dependence of polarization intensity and EVPA with the MOG parameter $\alpha$ vary periodically with the azimuthal coordinate $\phi$. For the fixed $\theta_o=20^{\circ}$ and $\chi=-90^{\circ}$, as $\phi$ varies in the range $(0,2\pi)$,  the polarization intensity and EVPA firstly decrease with $\alpha$, then increase and finally decrease once again. Similarly, for the different spin parameter $a$, the dependence of polarization intensity and EVPA on $\alpha$ is also qualitatively similar in this case. With the increase of $\chi$, the region of the polarization intensity increasing with $\alpha$ increases, but the region for the EVPA increasing with $\alpha$ decreases. As the observer inclination angle $\theta_o$ increases, the polarization intensity gradually becomes a monotonically increasing function of $\alpha$, but the change tendency of EVPA with $\alpha$ is similar to that in the case of the low inclination angle $\theta_o$. Moreover, the region where EVPA increases with $\alpha$ becomes narrow in the high inclination angle case.
\begin{figure}[htb!]
\includegraphics[width=14cm]{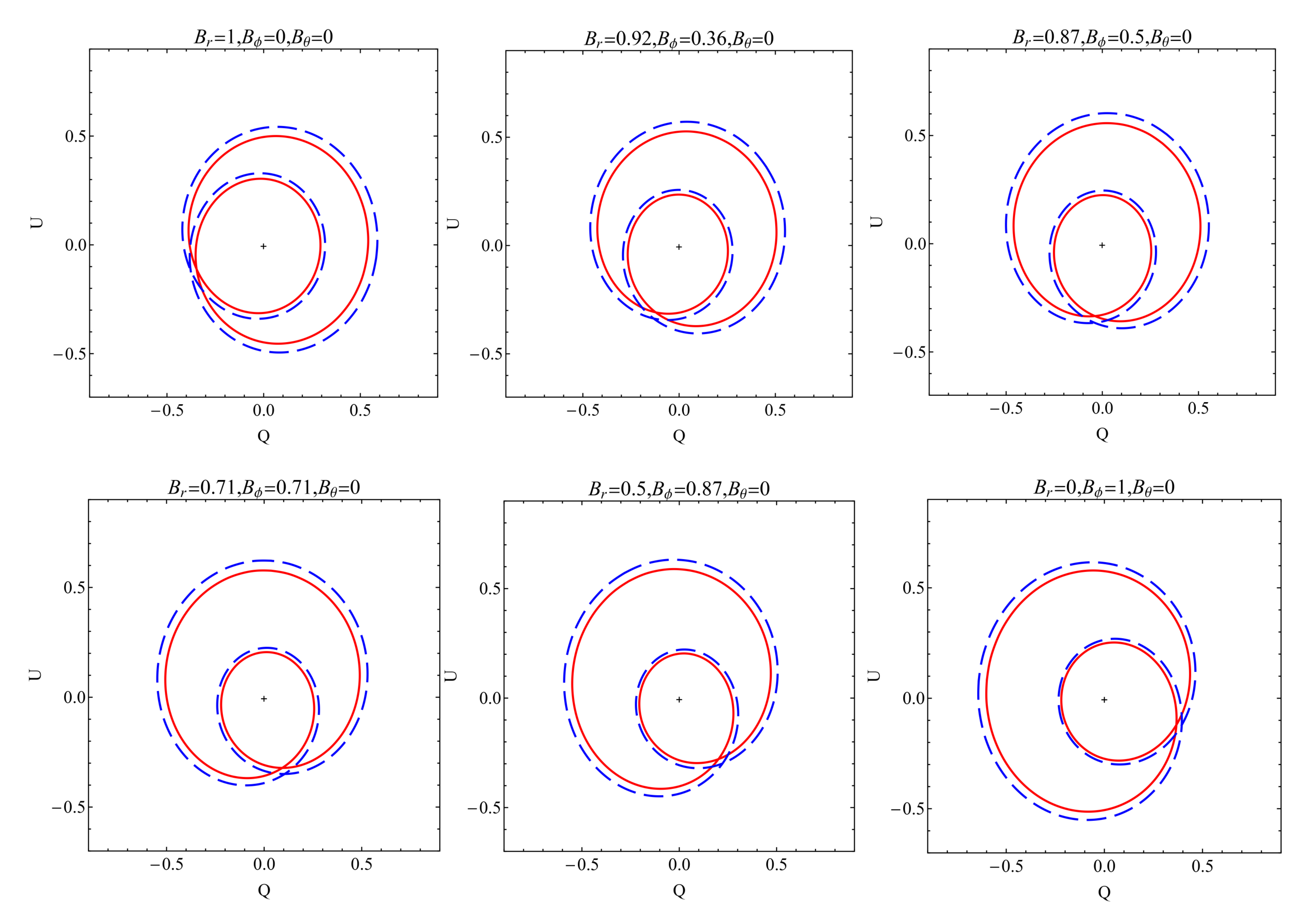}
\caption{Effects of the MOG parameter $\alpha$ on the $Q-U$ diagram for different equatorial magnetic fields in the Kerr-MOG black hole \eqref{Metric01}. Here $r_s=6$, $\theta_o=20^{\circ}$, $a=-0.3$, $\beta_\nu=0.3$, and $\chi=-90^{\circ}$. The blue dashed line and the red solid line correspond to the cases with the MOG parameter $\alpha=9$ and $\alpha=0$, respectively. Black crosshairs indicate the origin of each plot.}
\label{f9}
\end{figure}
\begin{figure}[htb!]
\includegraphics[width=13cm]{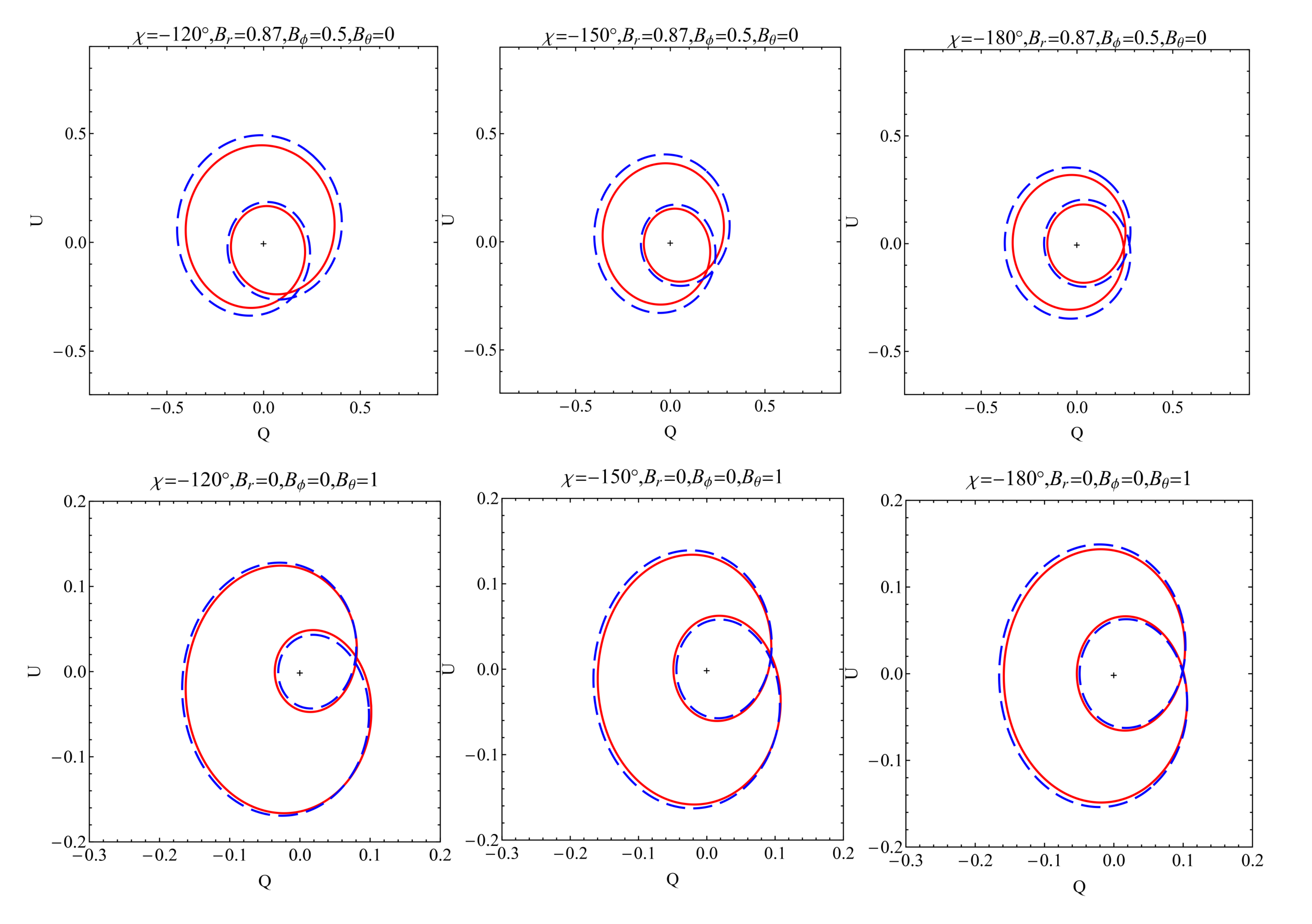}
\caption{Effects of the MOG parameter $\alpha$ on the $Q-U$ diagram for different fluid velocity angle $\chi$ in the Kerr-MOG black hole \eqref{Metric01}. Here $r_s=6$, $\theta_o=20^{\circ}$, $a=-0.3$, and $\beta_\nu=0.3$. The top and bottom rows correspond to the equatorial  magnetic field and the vertical one, respectively. The blue dashed line and the red solid line correspond to the cases with the MOG parameter $\alpha=9$ and $\alpha=0$, respectively. Black crosshairs indicate the origin of each plot.}
\label{f10}
\end{figure}
\begin{figure}[htb!]
\includegraphics[width=14cm]{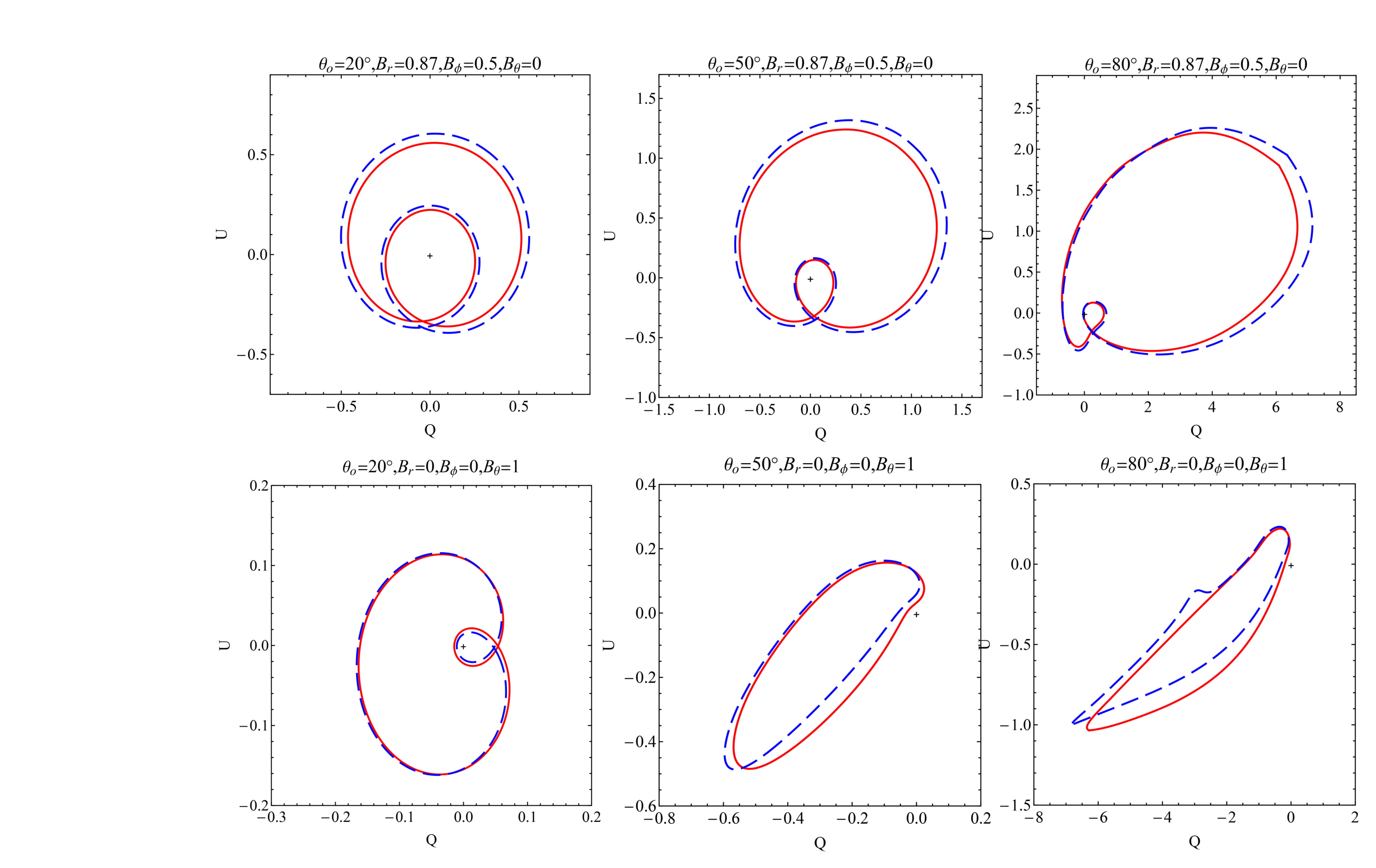}
\caption{Effects of the MOG parameter $\alpha$ on the $Q-U$ diagram for different observation inclination angle $\theta_o$ in the Kerr-MOG black hole \eqref{Metric01}. Here $r_s=6$, $a=-0.3$, $\beta_\nu=0.3$ and $\chi=-90^{\circ}$. The top and bottom rows represent the equatorial and vertical magnetic field, respectively. The blue dashed line and the red solid line correspond to the cases with the MOG parameter  $\alpha=9$ and $\alpha=0$, respectively. Black crosshairs indicate the origin of each plot.}
\label{f11}
\end{figure}
\begin{figure}[htb!]
\includegraphics[width=14cm]{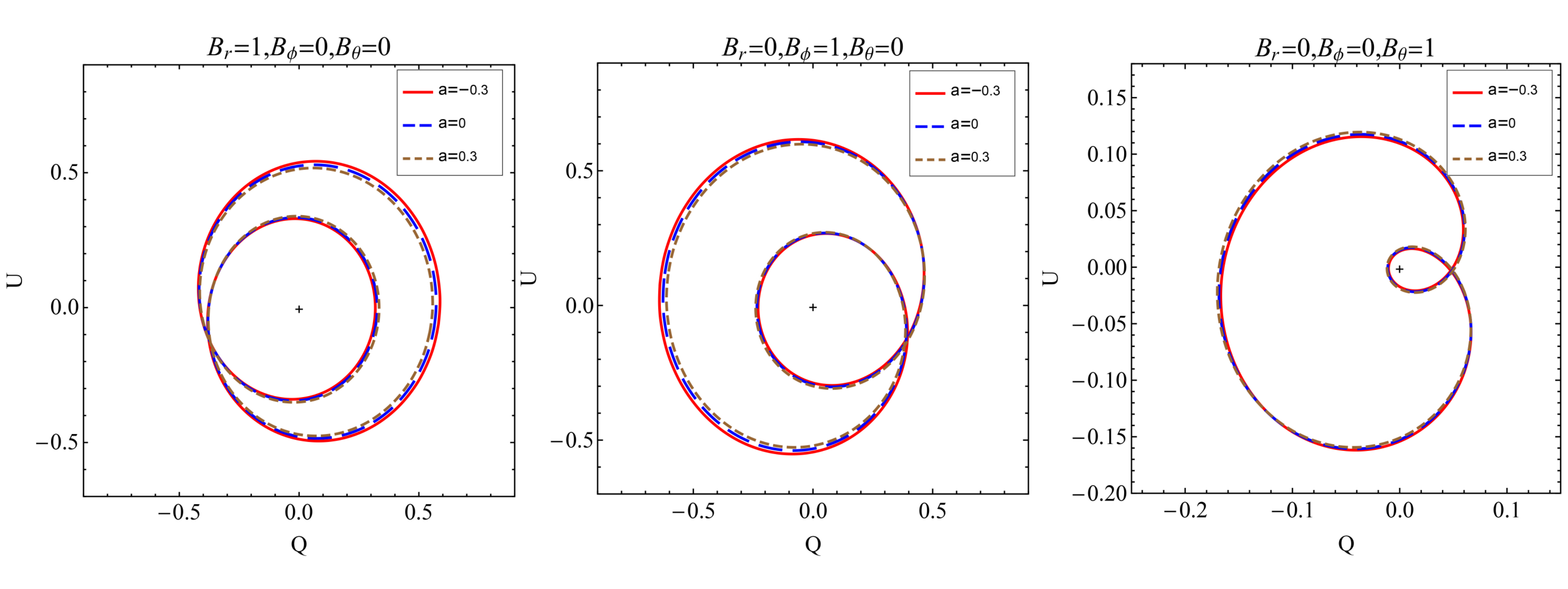}
\caption{Effects of the spin parameter $a$  on the $Q-U$ diagram for different magnetic fields in the Kerr-MOG black hole \eqref{Metric01}. Here $r_s=6$, $\theta_o=20^{\circ}$, $\alpha=9$, $\beta_\nu=0.3$ and $\chi=-90^{\circ}$. Black crosshairs indicate the origin of each plot.}
\label{f12}
\end{figure}

Figs.(\ref{f9})-(\ref{f11}) show the effects of the MOG parameter $\alpha$ on Stokes $Q-U$ loop patterns in the image of the emitting ring. In general, there are two loops enclosing the origin in the $Q-U$ plane.  Effects of the MOG parameter $\alpha$ on the $Q-U$ diagram depend heavily on the magnetic field configuration, the fluid velocity, the observation inclination angle and the spin parameter of black hole. As the magnetic field lies in the equatorial plane, the observed two Stokes $Q-U$ loops  increase with the MOG parameter $\alpha$ for the observer with $\theta_o=20^{\circ}$ and different fluid direction angle $\chi$.  For the high observation inclination angle, the sizes of two loops increases with $\alpha$ although the inner loop dramatically shrinks. As the magnetic field is vertical to the equatorial plane, the size of the outer loop increases  and the inner loop decreases with $\alpha$ for the observer with $\theta_o=20^{\circ}$ and different fluid direction angle $\chi$. For the high observation inclination angle, the inner loop vanishes and  the change of loop size with the MOG parameter $\alpha$ becomes more complicated.
For the fixed $\alpha$, as the spin parameter increases, Fig.(\ref{f12}) shows that the inner loop increases for different magnetic fields. However, the outer loop decreases with $a$ as the magnetic field lies in the equatorial plane, and its no longer changes monotonously with $a$ in the case of with the vertical magnetic field. Moreover, for the fixed $\alpha$, the changes of the $Q-U$ loop patterns with the magnetic field configuration, the fluid velocity and the observation inclination angle are similar to those in the Kerr black hole case.

\begin{figure}[htb!]
\includegraphics[width=15cm]{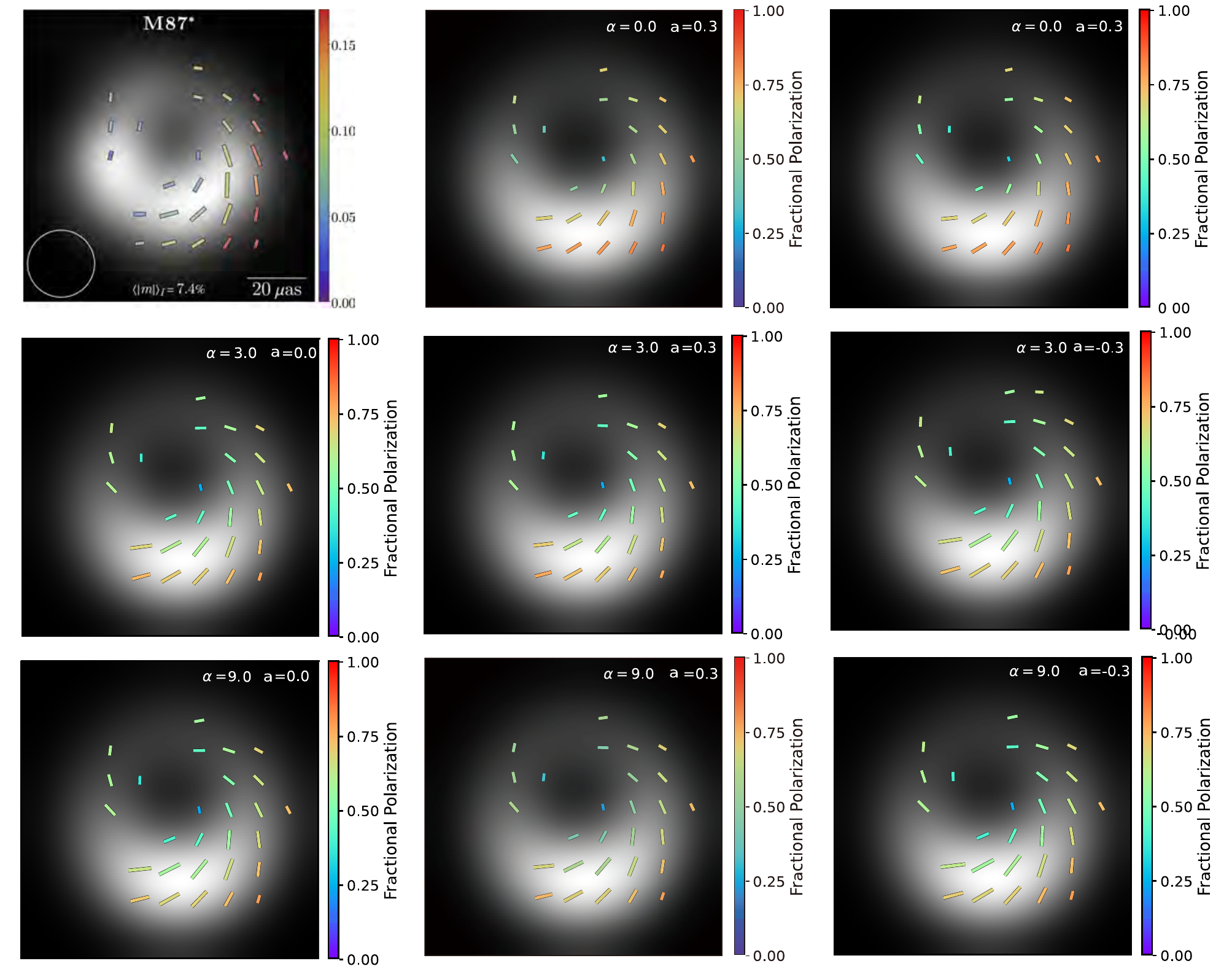}
\caption{Comparison between the polarimetric image of Kerr-MOG black hole and the black hole M 87* \cite{EHT8,EHT9}. Here $\theta=17^{\circ}$, $\beta_\nu=0.4$, $\chi=-150^{\circ}$, $B_r=0.87$, $B_\phi=0.5$ and $B_\theta=0$.}
\label{f13}
\end{figure}
Finally, Fig.(\ref{f13}) gives a comparison of the polarization image between a Kerr-MOG black hole and M87* black hole. Results show that the polarimetric images of Kerr-MOG black hole for different $\alpha$ own some spiral structures and their polarization patterns are similar to those of M87* observed in \cite{EHT8,EHT9}. The small difference in the polarization images in Fig.(\ref{f13}) implies that there remains some possibility for the STVG-MOG theory.

\section{Summary}

This study investigated the polarized images of the emitting ring around a Kerr-MOG black hole with an additional dimensionless parameter $\alpha$. The results show that for the fixed MOG parameter $\alpha$, the change of the polarization vector in the image plane with the spin parameter, the magnetic field configuration, the fluid velocity and the observation inclination angle are similar to those in the Kerr black hole case. The effects of the MOG parameter $\alpha$ on the observed polarization vector depend on the black hole parameters, the material distribution around black hole and the observation inclination angle. For the cases where the magnetic field lies in the equatorial plane, the polarization intensity increase with the MOG parameter $\alpha$ in the lower observation inclination angle case, and no longer varies monotonously
in the higher observation inclination angle case. The change of EVPA with $\alpha$ becomes more complicated and it also depends on the ratio between the radial component and the azimuthal component of magnetic field $B_r/B_\phi$. As the magnetic field is perpendicular to the equatorial plane,
the dependence of polarization intensity and EVPA with the MOG parameter $\alpha$ vary periodically with the azimuthal coordinate $\phi$ in the lower observation inclination angle case. As the observer inclination angle $\theta_o$ increases, the polarization intensity gradually becomes a monotonically increasing function of $\alpha$, but the change tendency of EVPA with $\alpha$ is similar to that in the case of the low inclination angle $\theta_o$ and the region where EVPA increases with $\alpha$ becomes more narrow.

Effects of the MOG parameter $\alpha$ on the $Q-U$ diagram also depend heavily on the magnetic field configuration, the fluid velocity, the observation inclination angle and the spin parameter of black hole. In the case with the lower observed inclination, the size of the $Q-U$ loop increases with $\alpha$ as the magnetic field lies in the equatorial plane, and  the size of the outer loop increases and the inner loop decreases as
the magnetic field is vertical to the equatorial plane. In the higher inclination angle case, the sizes of two loops increase with $\alpha$ although the inner loop dramatically shrinks as the magnetic field lies in the equatorial plane, however, as the magnetic field is vertical to the equatorial plane, the inner loop  vanishes and  the change of loop size with the MOG parameter $\alpha$ becomes more complicated. Moreover,
for the fixed $\alpha$, as the spin parameter increases,  the inner loop increases for different magnetic fields. However, the outer loop decreases with $a$ as the magnetic field lies in the equatorial plane, and its no longer changes monotonously with $a$ in the case of with the vertical magnetic field. Finally, making a comparison of the polarization images between Kerr-MOG black hole and M87*, there are some similar spiral structures appeared in the polarimetric images of two black holes. Our result also implies that there remains some possibility for the STVG-MOG theory.

\section{\bf Acknowledgments}

This work was  supported by the National Natural Science
Foundation of China under Grant No.11875026, 11875025 and 12035005.


\vspace*{0.2cm}

\begin{thebibliography}{99}
\baselineskip=0.5 cm
\bibitem{EHT1} The Event Horizon Telescope Collaboration, \textit{First M87 Event Horizon Telescope Results. I. The Shadow of the Supermassive Black Hole}, Astrophys. J. Lett. {\bf875}, L1 (2019).
\bibitem{EHT2} The Event Horizon Telescope Collaboration, \textit{First M87 Event Horizon Telescope Results. II. Array and Instrumentation}, Astrophys. J. Lett. {\bf875}, L2 (2019).
\bibitem{EHT3} The Event Horizon Telescope Collaboration, \textit{First M87 Event Horizon Telescope Results. III. Date Processing and Calibration}, Astrophys. J. Lett. {\bf875}, L3 (2019).
\bibitem{EHT4} The Event Horizon Telescope Collaboration, \textit{First M87 Event Horizon Telescope Results. IV. Imaging the Central Supermassive Black Hole}, Astrophys. J. Lett. {\bf875}, L4 (2019).
\bibitem{EHT5} The Event Horizon Telescope Collaboration, \textit{First M87 Event Horizon Telescope Results. V. Physical origin of the asymmetric ring}, Astrophys. J. Lett. {\bf875}, L5 (2019).
\bibitem{EHT6} The Event Horizon Telescope Collaboration, \textit{First M87 Event Horizon Telescope Results. VI. The Shadow and Mass of the Central Black Hole}, Astrophys. J. Lett. {\bf875}, L6 (2019).
\bibitem{EHT7} The Event Horizon Telescope Collaboration, \textit{First Sagittarius $A^\ast$ Event Horizon Telescope Results. I. The Shadow of the Supermassive Black Hole in the Center of the Milky Way}, Astrophys. J. Lett. {\bf930}, L12 (2022).
\bibitem{EHT8} The Event Horizon Telescope Collaboration, \textit{First M87 Event Horizon Telescope Results. VII. Polarization of the Ring}, Astrophys. J. Lett. {\bf875}, L7 (2019).
\bibitem{EHT9} The Event Horizon Telescope Collaboration, \textit{First M87 Event Horizon Telescope Results. VIII. Magnetic Field Structure near The Event Horizon}, Astrophys. J. Lett. {\bf875}, L8 (2019).

\bibitem{PZ1} R. Narayan, D. C. M. Palumbo, M. D. Johnson, Z. Gelles, E. Himwich, D. O. Chang, A. Ricarte, J. Dexter, C. F. Gammie, A. A. Chael, and The Event
Horizon Telescope Collaboration, \textit{The Polarized Image of a Synchrotron-emitting Ring of Gas Orbiting a Black Hole}, Astrophys. J. {\bf912}, 35 (2021).
\bibitem{PZ2} Z. Gelles, E. Himwich, D. C. M. Palumbo, M. D. Johnson, \textit{Polarized Image of Equatorial Emission in the Kerr Geometry}, Physical Review D 104, (2021) 044060. arXiv:2105.09440.
\bibitem{PZ3} X. Qin, S. Chen, J. Jing, \textit{Polarized image of an equatorial emitting ring around a 4D Gauss-Bonnet black hole}, arXiv: 2111.10138.
\bibitem{PZ4} X. Liu, S. Chen, J. Jing, \textit{Polarization distribution in the image of a synchrotron emitting ring around regular black holes}, arXiv: 2205.00391.
\bibitem{PZ5} Z. Zhang, S. Chen, X. Qin and J. Jing, \textit{Polarized image of a Schwarzschild black hole with a thin accretion disk as photon couples to Weyl tensor}, Eur. Phys. J. C. {\bf81},  11 (2021) 991. arXiv: 2106.07981. (2021).
\bibitem{PZ6} Z. Zhang, S. Chen, and J. Jing, \textit{Image of Bonnor black dihole with a thin accretion disk and its polarization information}, arXiv: 2205.13396 (2021).
\bibitem{PZ7} H. Zhu and M. Guo, \textit{Polarized image of synchrotron radiations of hotspots in Schwarzschilld-Melvin black hole spacetime}, arXiv: 2205.04777.
\bibitem{PZ8} Z. Hu, Y. Hou, H. Yan, M. Guo, and B. Chen, \textit{Electromagnetic radiations and polarized images of synchrotron radiations in curved spacetime}, arXiv: 2203.02908.


\bibitem{PZJG1} P. A. Connors, T. Piran, and R. F. Stark, \textit{Polarization features of X-ray radiation emitted near black holes}, Astrophys. J. {\bf235}, 224-244 (1980).
\bibitem{PZJG2} B. C. Bromley, F. Melia, and S. Liu, \textit{Polarimetric Imaging of the Massive Black Hole at the Galactic Center}, Astrophys. J. Lett. {\bf555}, L83-L86 (2001).
\bibitem{PZJG3} L.X. Li, R. Narayan, and J. E. McClintock, \textit{Inferring the Inclination of a Black Hole Accretion Disk from Observations of its Polarized Continuum Radiation}, Astrophys. J. {\bf555} no. 1, 847-865 (2009).
\bibitem{PZJG4} R. V. Shcherbakov, R. F. Penna, and J. C. McKinney, \textit{Sagittarius $A^\ast$ Accretion Flow and Black Hole Parameters from General Relativistic Dynamical and Polarized Radiative Modeling}, Astrophys. J. {\bf755}, 847-865 (2012).
\bibitem{PZJG5} J. Dexter, \textit{A public code for general relativistic, polarised radiative transfer around spinning black holes}, MNRAS. {\bf462} no. 1, 115-136 (2016).
\bibitem{PZJG6} R. Gold, J. C. McKinney, M. D. Johnson, and S. S. Doeleman, \textit{Probing the Magnetic Field Structure in Sgr $A^\star$ on Black Hole Horizon Scales with Polarized Radiative Transfer Simulations}, Astrophys. J. {\bf837}, 180 (2017).
\bibitem{PZJG7} F. Marin, M. Dov\u{c}iak, F. Muleri, F. F. Kislat, and H. S. Krawczynski, \textit{Predicting the X-ray polarization of type 2 Seyfert galaxies}, MNRAS. {\bf473}, 1286-1316 (2016).
\bibitem{PZJG8} A. Jim'enez-Rosales and J. Dexter, \textit{The impact of Faraday effects on polarized black hole images of Sagittarius $A^\star$}, MNRAS. {\bf478}, 1875-1883 (2018).
\bibitem{PZJG9} D. C. M. Palumbo, G. N. Wong, and B. S. Prather, \textit{Discriminating accretion states via rotational symmetry in simulated polarimetric images of m87}, Astrophys. J. {\bf894}, 156 (2020).
\bibitem{PZJG10} M. Mo¡äscibrodzka, \textit{General relativistic polarized radiative transfer with inverse-Compton scatterings}, MNRAS. {\bf491} no. 4, 4807-4815 (2020).
\bibitem{PZJG11} M. Moscibrodzka, A. Janiuk, and M. De Laurentis, \textit{Unraveling circular polarimetric images of magnetically arrested accretion flows near event horizon of a black hole}, arXiv: 2103.00267. (2021)

\bibitem{obs1} V. C. Rubin, E. M. Burbidge, G. R. Burbidge,  K. H. Prendergast, \textit{ The Rotation and Mass of the Inner Parts of NGC 4826}, Astrophys. J. {\bf141}, 885 (1965).
\bibitem{obs2} V. C. Rubin, W. K. Ford, \textit{Rotation of the Andromeda Nebula from a Spectroscopic Survey of Emission Regions}, Astrophys. J. {\bf159}, 379 (1970).


\bibitem{MOG1} J. W. Moffat, \textit{Scalar-Tensor-Vector Gravity Theory}, J. Cosmol. Astropart. Phys. {\bf2006}(3), 004 (2006).

\bibitem{MOG4} J. W. Moffat, S. Rahvar, \textit{The MOG weak field approximation and observational test of galaxy rotation curves}, Mon. Not. Roy. Astron. Soc. {\bf436}, 1439 (2013).
 \bibitem{MOG41s}J. W. Moffat and V. T. Toth, \textit{Rotational Velocity Curves in the Milky Way as a Test of Modified Gravity}, Phys. Rev. D {\bf91}, 043004 (2015), arXiv:1411.6701 [astro-ph.GA].
 \bibitem{MOG2} J. W. Moffat, \textit{Gravitational Lensing in Modified Gravity and the Lensing of Merging Clusters without Dark Matter}, astro-ph/0608675.
\bibitem{MOG3} J. R. Brownstein, J. W. Moffat, \textit{The Bullet Cluster 1E0657-558 evidence shows Modified Gravity in the absence of Dark Matter}, Mon. Not. Roy. Astron. Soc. {\bf382}, 29 (2007).
\bibitem{MOG5} J. W. Moffat, S. Rahvar, \textit{The MOG Weak Field approximation II. Observational test of Chandra X-ray Clusters}, Mon. Not. Roy. Astron. Soc. {\bf441}, 3724 (2014).

\bibitem{MOG6} J. W. Moffat, \textit{LIGO GW150914 and GW151226 Gravitational Wave Detection and Generalized Gravitation Theory (MOG)}, Phys. Lett. B {\bf763}, 427 (2016).
\bibitem{MOG7} M. A. Green, J. W. Moffat, V. T. Toth, \textit{Modified Gravity (MOG), the speed of gravitational radiation and the event GW170817/GRB170817A}, Phys. Lett. B {\bf780}, 300 (2018).
\bibitem{MOG8} J. W. Moffat, \textit{Black holes in modified gravity (MOG)}, Eur. Phys. J. C {\bf75}, 175 (2015). arXiv:1412.5424.
\bibitem{MOG17} L. Manfredi, J. Mureika, and J. W. Moffat, \textit{Quasinormal modes of modified gravity (MOG) black holes}, Phys. Lett. B {\bf779}, 492 (2018). arXiv:1711.03199.
\bibitem{MOG9} J. W. Moffat, \textit{Modified gravity black holes and their observable shadows}, Eur. Phys. J. C {\bf75}, 130 (2015). arXiv:1502.01677.
\bibitem{MOG19} H. Wang, Y. Xu and S. Wei, \textit{Shadows of Kerr-like black holes in a modified gravity theory}, J. Cosmol. Astropart. Phys. {\bf1903}, 046 (2019). arXiv:1810.12767.
\bibitem{MOG10} J. R. Mureika, J. W. Moffat, and M. Faizal, \textit{Black hole thermodynamics in modified gravity (MOG)}, Phys. Lett. B {\bf757}, 528 (2016). arXiv:1504.08226.

\bibitem{MOG11} H. C. Lee and Y. J. Han, \textit{Inner-most stable circular orbit in Kerr-MOG black hole}, Eur. Phys. J. C {\bf77}, 655 (2017). arXiv:1704.02740.
\bibitem{MOG12} M. Sharif and M. Shahzadi, \textit{Particle dynamics near Kerr-MOG black hole}, Eur. Phys. J. C {\bf77}, 363 (2017). arXiv:1705.03058.
\bibitem{MOG13} S. Hussain and M. Jamil, \textit{Timelike geodesics of a modified gravity black hole immersed in an axially symmetric magnetic field}, Phys. Rev. D {\bf92}, 043008 (2015). arXiv:1508.02123.
\bibitem{MOG14} D. Perez, F. G. L. Armengol, and G. E. Romero, \textit{Accretion disks around black holes in Scalar-Tensor-Vector gravity}, Phys. Rev. D {\bf95}, 104047 (2017). arXiv:1705.02713.
\bibitem{MOG15} P. Sheoran, A. Herrera-Aguilar, and U. Nucamendi, \textit{Mass and spin of a Kerr black hole in modified gravity and a test of the Kerr black hole hypothesis}, Phys. Rev. D {\bf97}, 124049 (2018). arXiv:1712.03344.
\bibitem{MOG16} J. W. Moffat, \textit{Misaligned spin merging black holes in modified gravity (MOG)}, arXiv:1706.05035.

\bibitem{MOG18} S. Wei, Y. Liu, \textit{Merger estimates for rotating Kerr black holes in modified gravity}, Phys. Rev. D 98, 024042 (2018). arXiv:1803.09530.

\bibitem{Math1} S. E. Gralla and A. Lupsasca, \textit{Lensing by Kerr black holes}, Physical Review D {\bf101}, 044031 (2020).  arXiv:1910.12873.
\bibitem{PWconstant} M. Walker and R. Penrose, \textit{On quadratic first integrals of the geodesic equations for type $\{22\}$ spacetimes}, Commun. Math. Phys. {\bf18}, 265 (2001).
\bibitem{Chandrasekhar} S. Chandrasekhar, \textit{The mathematical theory of black holes}. 1985



\bibitem{Himwich} E. Himwich, M. D. Johnson, A. Lupsasca, and A. Strominger, \textit{Universal polarimetric signatures of the black hole photon ring}, Phys. Rev. D. {\bf101}, 084020 (2020).



\end{thebibliography}
\end{document}